\documentstyle[aps]{revtex}
\input epsf.tex
\begin{document}
\title{Perturbations in a coupled scalar field cosmology}
\author{Luca Amendola}
\address{Osservatorio Astronomico di Roma, \\
Viale Parco Mellini 84, \\
00136 Roma, Italy\\
{\it amendola@oarhp1.rm.astro.it}}
\date{\today }
\maketitle

\begin{abstract}
I analyze the density perturbations in a cosmological model with a scalar
field coupled to ordinary matter, such as one obtains in string theory and
in conformally transformed scalar-tensor theories. The spectrum of
multipoles on the last scattering surface and the power spectrum at the
present are compared with observations to derive bounds on the coupling
constant and on the exponential potential slope. It is found that the
acoustic peaks and the power spectrum are strongly sensitive to the model
parameters. The models that best fit the galaxy spectrum and satisfy the
cluster abundance normalization have field energy density $\Omega _{\phi
}\simeq 0.05-0.15$ and a scale factor expansion law $a\sim t^{p}$, $p\simeq
0.67-0.69.$
\end{abstract}

\section{Introduction}

\bigskip

Perhaps the most important concept in modern cosmology is that fundamental
physics, along with gravity, shapes the distribution of matter on very large
scales. Fundamental physics enters in at least two distinct ways: through
the potential of the inflationary field, which sets the initial conditions
of the fluctuation field, and through the properties of the dark matter,
which govern the evolution of the fluctuations up to the present. As a
consequence, the imprint of the density fluctuations on the microwave
background and on the galaxy distribution allows tests of basic laws of
physics that, in many cases, could not be realized with any other mean.

An impressive array of different proposals have been formulated for as
concerns the inflationary side of the story, that is, the initial
conditions. So far, there is not an overwhelming reason to modify the
simplest inflationary prescription of a flat spectrum, although several
variations on the theme, like a small tilting (Lucchin \& Matarrese 1985,
Cen et al. 1992) or some break in the scale invariance (Gottloeber et al.
1994, Amendola et al. 1995) or the contribute of primordial voids (Amendola
et al. 1996) cannot be excluded either.

Similarly, many theories have been proposed for as concerns the evolution of
the fluctuations, trying to elucidate the nature and properties of the dark
matter component. A partial list of the dark matter recipes includes the
standard CDM and variations such as CDM plus a hot component (MDM), or plus
a cosmological constant ($\Lambda $CDM), or plus a scalar field ($\phi $
CDM). The latter class of models, in particular, has been explored greatly
in recent times, to various purposes. First, a light scalar field is
predicted by many fundamental theories (string theory,
pseudo-Nambu-Goldstone model, Brans-Dicke theory etc), so that it is natural
to look at its cosmological consequences (Wetterich 1995, Frieman et al.
1995, Ferreira \& Joyce 1998). Second, a scalar field may produce an
effective cosmological constant, with the benefit that its dynamics can be
linked to some underlying theory, or can help escape the strong constraints
on a true cosmological constant (Coble, Dodelson \& Frieman 1997, Waga \&
Miceli 1999). In turn, this effective cosmological constant can be tuned to
explain the observation of an accelerated expansion (Perlmutter et al. 1998,
Riess et al. 1998) and to fix the standard CDM spectrum as well (Zlatev et
al. 1998, Caldwell et al. 1998, Perrotta \& Baccigalupi 1999, Viana \&
Liddle 1998). Finally, even a small amount of scalar field density may give
a detectable contribution to the standard CDM scenario, similar to what one
has in the MDM model (Ferreira \& Joyce 1998, hereinafter FJ).

In this paper we pursue the investigation of the effects of a scalar field
in cosmology by adding a coupling between the field and ordinary matter.
Such a coupling has been proposed and studied several times in the past
(e.g. Ellis et al. 1989, Wetterich 1995) but, as far as we know, its
consequences on the microwave background and on the power spectrum have not
been determined. Scope of this paper is to solve the fluctuation equations
for a coupled scalar field theory, and to compare with the already available
data in the microwave sky and in galaxy surveys. We refer to this model as
coupled $\phi $CDM. Up to a conformal transformation, the model we study is
equivalent to a non-minimal coupling theory in which the scalar field
couples to gravity like in a Brans-Dicke Lagrangian; the perturbations in
such models have been investigated by Chen \& Kamionkowsky (1999) and
Baccigalupi et al. (1999) in a background in which the scalar field acts
like a dynamical cosmological constant (see also Uzan 1999). The model we
present here is in fact more general, since a wide class of non-minimal
coupling models can be recast in the form we study below (Amendola et al.
1993, Wetterich 1995, Amendola 1999).

There are several models of cosmological scalar field in the literature,
essentially characterized by the scalar field potential and by the initial
conditions. We can divide the models in two broad class: in the first one,
the field potential energy dominates at the present, so that it resembles a
cosmological constant. In the second one, the field kinetic energy is not
negligible, and the field adds to the ordinary matter as an additional
component, like in MDM models. To this second class belongs the model of FJ.
They adopt an exponential potential for the scalar field, able to drive an
attractor scaling solution which self-adjusts to the dominant matter
component. In such a model, the density fraction of the field does not
depends on the initial conditions, but is determined by the potential
parameters. Therefore, the coincidence that the energy density in the field
and in the matter components are comparable can be explained by the
underlying physics (the field potential) rather than by the initial
conditions. Although the coupling we will introduce can be applied to any
scalar field model, we focus our attention in this paper on the exponential
potential model of FJ. Beside being particularly simple, because of its
attractor properties (Wetterich 1988, Ratra \& Peebles 1988), such a model
is also easily falsifiable, because the effect of the scalar field is
important at all times (not just at the present as when the field acts as a
cosmological constant), and therefore induces a strong effect on the
cosmological sky. As a consequence, the constraints we derive on the model
parameters are rather strong.

The same exponential potential also allows solutions which belong to the
first class mentioned above, in which the field acts much like a
cosmological constant, and drives at the present an accelerated expansion.
These solutions, and their linear perturbations, have been studied by Viana
\& Liddle (1998) and Caldwell et al. (1998). The effect of adding a coupling
to these models will be analyzed in a subsequent work.

\section{Coupled scalar field model}

Consider two components, a scalar field $\phi $ and ordinary matter (e.g.,
baryons plus CDM) described by the energy-momentum tensors $T_{\mu \nu (\phi
)}$ and $T_{\mu \nu (m)}$. General covariance requires the conservation of
their sum, so that it is possible to consider a coupling such that, for
instance, 
\begin{eqnarray}
T_{\nu (\phi );\mu }^{\mu } &=&CT_{(m)}\phi _{;\mu },  \nonumber \\
T_{\nu (m);\mu }^{\mu } &=&-CT_{(m)}\phi _{;\mu }.  \label{coup1}
\end{eqnarray}
Such a coupling arises for instance in string theory, or after a conformal
transformation of Brans-Dicke theory (Wetterich 1995, Amendola 1999). It has
also been proposed to explain 'fifth-force' experiments, since it
corresponds to a new interaction that can compete with gravity and be
material-dependent. The coupling arises from Lagrangian terms of the form
(Wetterich 1995) 
\begin{equation}
-m_{\psi }^{2}\exp (-C\kappa \phi )\psi _{,\mu }\psi ^{,\mu },
\label{string}
\end{equation}
where $\kappa ^{2}=8\pi G$ and $\psi $ is the ordinary matter field of mass $ 
m_{\psi }$, e.g., the nucleon field.

The specific coupling (\ref{coup1}) is only one of the possible form.
Non-linear couplings as $CT_{(m)}F(\phi )\phi _{;\mu }$ or more complicate
functions are also possible. Also, one can think of different coupling to
different matter species, for instance coupling the scalar field only to
dark matter and not to baryons. Such a species-dependent coupling has been
proposed by Damour, Gibbons \& Gundlach (1990), \ and shown to be
observationally viable. Notice that the coupling to radiation (subscript $ 
\gamma $) vanishes, since $T_{(\gamma )}=0.$ Here we restrict ourselves to
the simplest possibility (\ref{coup1}), which is also the same investigated
earlier by Wetterich (1995) and is the kind of coupling that arises from
Brans-Dicke models. For instance, a field with coupling to gravity $\frac{1}{ 
2}\xi \phi ^{2}R$ in the Lagrangian acquires, after conformal
transformation, a coupling to matter of the form (\ref{coup1}) with 
\begin{equation}
C=\frac{\kappa }{\left( 6+1/\xi \right) ^{1/2}};
\end{equation}
in the limit of small positive coupling this becomes 
\begin{equation}
C=\kappa \sqrt{\xi }.
\end{equation}

There are several constraints on the coupling constant $C$ along with
constraints on the mass of the scalar field particles, reviewed by Ellis et
al. (1989). The constraints arise from a variety of observations, ranging
from Cavendish-type experiments, to primordial nucleosynthesis bounds, to
stellar structure, etc etc. Most of them, however, apply only if the scalar
couples to baryons, which is not necessarily the case, and/or involve the
mass of the scalar field particles, which is unknown. The most stringent
bound, quoted by Wetterich (1995) amounts to 
\begin{equation}
|C|<0.1M_{P}^{-1}  \label{wetlimit}
\end{equation}
but again holds only for a coupling to baryons. Moreover, these constraints
are local both in space and time, and could be easily escaped by a
time-dependent coupling constant. In the following we leave therefore $C$ as
a free parameter.

The constraints from nucleosynthesis refer to the energy density in the
scalar component. This has to be small enough not to perturb element
production, so that at the epoch of nucleosynthesis (Wetterich 1995, Sarkar
1996, FJ) 
\begin{equation}
\Omega _{\phi }<0.1-0.2
\end{equation}
We will see that this bound is satisfied by all the interesting models.

There is an immediate consequence of the coupling for as concerns cosmology.
The coupling modifies the conservation equation for the ordinary matter,
leading to a different effective equation of state for the matter. This
alters the scale factor expansion law in matter dominated era (MDE) from $ 
a\sim t^{2/3}$ to $t^{p}$, $p\neq 2/3$ . In turn, this has three effects.
First, the sound horizon at decoupling (when decoupling occurs in MDE) is
modified with respect to the uncoupled case, being larger for $p>2/3$ and
smaller in the opposite case, as we will show. This modifies the peak
structure of the microwave background multipoles. Second, the epoch of
matter/radiation equivalence moves to a later epoch if $p>2/3$ and to an
earlier epoch in the opposite case. This shifts the range of scales for
which there is the growth suppression of the sub-horizon modes in radiation
dominated era (RDE), leading to a turnaround of the power spectrum on larger
scales if $p>2/3$ (smaller if $p<2/3$). Finally, the different scale factor
law modifies the fluctuation growth for sub-horizon modes in the MDE,
generally reducing the growth for all values of $C$. Similar effects have
been observed by Chen \& Kamionkowsky (1999) and Baccigalupi et al. (1999)
in Brans-Dicke models. The next sections investigate these effects in detail.

\section{Background}

Here we derive the background equations in the conformal FRW metric 
\begin{equation}
ds^{2}=a^{2}(-d\tau ^{2}+\delta _{ij}dx^{i}dx^{j}).
\end{equation}
The scalar field equation is 
\begin{equation}
\ddot{\phi}+2H\dot{\phi}+a^{2}U_{,\phi }=C\rho _{m}a^{2},
\end{equation}
where $H=\dot{a}/a$ , and we adopt the exponential potential 
\begin{equation}
U(\phi )=Ae^{s\phi }.
\end{equation}
The matter (subscript $m$) and the radiation (subscript $\gamma $) equations
are 
\begin{eqnarray}
\dot{\rho}_{m}+3H\rho _{m} &=&-C\rho _{m}\dot{\phi} \\
\dot{\rho}_{\gamma }+4H\rho _{\gamma } &=&0.
\end{eqnarray}
Denoting with $\tau _{0}$ the conformal time today, let us put 
\begin{equation}
a(\tau _{0})=1,\quad \rho _{m}(\tau _{0})=\frac{3H_{0}^{2}}{8\pi }\Omega
_{m}=\rho _{m0},\quad \rho _{\gamma }(\tau _{0})=\rho _{\gamma 0},\quad \phi
(\tau _{0})=\phi _{0}.
\end{equation}
Without loss of generality, the scalar field can be rescaled by a constant
quantity, by a suitable redefinition of the potential constant $A$. We put
then $\phi _{0}=0$. This gives 
\begin{eqnarray}
\rho _{m} &=&\rho _{m0}a^{-3}e^{-C\phi },  \nonumber \\
\rho _{\gamma } &=&\rho _{\gamma 0}a^{-4}.
\end{eqnarray}
The (0,0) Einstein equation can be written 
\begin{equation}
H^{2}=\frac{\kappa ^{2}}{3}\left( \frac{\rho _{m0}}{a}e^{-C\phi }+\frac{\rho
_{\gamma 0}}{a^{2}}+\frac{1}{2}\dot{\phi}^{2}+Ua^{2}\right) .
\end{equation}

The dynamics of the model is very simple to study in the regime in which
either matter or radiation dominates. Assume that the dominant component has
equation of state 
\begin{equation}
p=(w-1)\rho .
\end{equation}
Then, following Copeland et al. (1997) we define 
\begin{equation}
x=\frac{\kappa \dot{\phi}}{\sqrt{6}H},\quad y=\frac{\kappa \sqrt{U}}{\sqrt{3} 
H},
\end{equation}
and introduce the independent variable $\alpha =\log a$. Notice that $x^{2}$
and $y^{2}$ give the fraction of total energy density carried by the scalar
field kinetic and potential energy, respectively . Then, we can rewrite the
equations as (Amendola 1999) 
\begin{eqnarray}
x^{\prime } &=&-3x+3x\left[ x^{2}+\frac{1}{2}w(1-x^{2}-y^{2})\right] -\mu
y^{2}+\beta (1-x^{2}-y^{2}),  \nonumber \\
y^{\prime } &=&\mu xy+3y\left[ x^{2}+\frac{1}{2}w(1-x^{2}-y^{2})\right] .
\label{syst2}
\end{eqnarray}
where the prime denotes here $d/d\alpha $ and where we introduce the
adimensional constants 
\begin{equation}
\beta =\sqrt{\frac{3}{2}}\frac{C}{\kappa },\quad \mu =\sqrt{\frac{3}{2}} 
\frac{s}{\kappa },\quad 
\end{equation}
(in Amendola 1999 we defined $\beta $ one half of the definition above).
Notice that in this simplified system with a single component, plus the
scalar field, the constant $\beta $ is the coupling constant for the
dominant component only, so that we are implicitly assuming $\beta =0$
during RDE. The system is invariant under the change of sign of $y$ and of $ 
\alpha $. Since it is also limited by the condition $\rho >0$ to the circle $ 
x^{2}+y^{2}\leq 1$, we limit the analysis only to the unitary semicircle of
positive $y$. The critical points, i.e. the points that verify $x^{\prime
}=y^{\prime }=0$, are scaling solutions, on which the scalar field equation
of state is 
\begin{equation}
w_{\phi }=\frac{2x^{2}}{x^{2}+y^{2}}=const,
\end{equation}
the scalar field total energy density is $\Omega _{\phi }=x^{2}+y^{2}$, and
the scale factor is 
\begin{equation}
a\sim \tau ^{\frac{p}{1-p}}=t^{p},\quad p=\frac{2}{3}\left[ \frac{1}{ 
w+\Omega _{\phi }(w_{\phi }-w)}\right] 
\end{equation}
($t$ being the time defined by $dt=a(\tau )d\tau $).

The system (\ref{syst2}) with an exponential potential has up to five
critical points, that can be classified according to the dominant energy
density: one dominated by the scalar field total energy density, one in
which the fractions of energy density in the matter and in the field are
both non-zero, one in which the matter field and the field kinetic energy
are both non-zero, while the field potential energy vanishes, and finally
two dominated by the kinetic energy of the scalar field (at $x=\pm 1$) . 
The critical points are listed in Tab. I, where we put $g(\beta ,w,\mu
)\equiv 4\beta ^{2}+4\beta \mu +18w.$ For any value of the parameters there
is one and only one stable critical point (attractor). More details on the
phase space dynamics in Amendola (1999) and, for $\beta=0$, 
in Copeland et al. (1997). 
\[
\begin{tabular}{|c|c|c|c|c|c|}
\hline
& $x$ & $y$ & $\Omega _{\phi }$ & $p$ & $w_{\phi }$ \\ \hline
$a$ & $-\mu /3$ & $\left( 1-x_{a}^{2}\right) ^{1/2}$ & $1$ & $3/\mu ^{2}$ & $ 
2\mu ^{2}/9$ \\ \hline
$b$ & $-\frac{3w}{2\left( \mu +\beta \right) }$ & $-x_{b}\left( \frac{g}{
9w^{2}}-1\right) ^{1/2}$ & $\frac{g}{4\left( \beta +\mu \right) ^{2}}$ & $ 
\frac{2}{3w}\left( 1+\frac{\beta }{\mu }\right) $ & $\frac{18w^{2}}{g}$ \\ 
\hline
$c$ & $\frac{2}{3}\frac{\beta }{2-w}$ & $0$ & $\frac{4}{9}\left( \frac{\beta 
}{2-w}\right) ^{2}$ & $\frac{6(2-w)}{4\beta ^{2}+9(2-w)w}$ & $2$ \\ \hline
$d$ & $-1$ & $0$ & $1$ & $1/3$ & $2$ \\ \hline
$e$ & $+1$ & $0$ & $1$ & $1/3$ & 2 \\ \hline
\multicolumn{6}{|c|}{Tab. I} \\ \hline
\end{tabular}
\]

The perturbations on solutions converging toward the attractor $a$ have been
studied in Viana \& Liddle (1998) and in Caldwell et al. (1999) for zero
coupling. In this case the scalar field is starting to dominate today, and
mimics a cosmological constant. The case of interest here is instead the
solution $b$ in Tab. I, since this is the only critical point which allow a
partition of the energy between the scalar field and the matter and
(contrary to $c$) is stable also in the RDE, when $\beta =0$. The solution $ 
b $ is compatible with a $p$ larger or smaller than $2/3$. It exists and is
stable (that is, is an attractor) in the region delimited by $\mu <\mu _{-}$
and $\mu >\mu _{+}$ and the two branches of the curve 
\begin{equation}
\mu _{0}=-\frac{1}{4\beta }\left( 4\beta ^{2}+18w-9w^{2}\right) .
\end{equation}
The scale factor slope on the attractor is (Wetterich 1995) 
\begin{equation}
p=\frac{2}{3w}\left( 1+\frac{\beta }{\mu }\right) .
\end{equation}
and, if $w=1$, is inflationary for 
\begin{equation}
2\beta >\mu .
\end{equation}
The parametric space region in which the attractor exists is shown in Fig.
1. For any value of the parameters $\beta ,\mu $ there is a pair of
observables $\Omega _{\phi },p$. When radiation dominates, $\beta =0$, and
the scale factor is the usual RDE one, $p=1/2$. The mapping from $\beta ,\mu 
$ to $\Omega _{\phi },p$ is shown in the same Fig. 1: as one can see, to get
a large $p$ a large $\Omega _{\phi }$ is also needed. In Fig. 2 we show the
phase space of the system for $\Omega _{\phi }=0.1,p=0.7$ assuming matter
domination. Notice that only for $\beta \neq 0$ there is the possibility to
get an inflationary attractor with $\Omega _{\phi }<1$, as some observations
suggest. It can be easily demonstrated that the coupled exponential
potential with $2\beta >\mu $ is the only model that allows inflationary
attractors with a non-vanishing matter component. Although such a
possibility is intriguing, it is hardly realistic, since an inflationary
expansion that lasted for most of the MDE would not allow any fluctuation
growth via gravitational instability.

When both radiation and matter are present, the system goes rapidly from the
radiation attractor, for which 
\begin{equation}
\quad \Omega _{\phi (R)}=6/\mu ^{2},\quad p_{R}=1/2,
\end{equation}
to the matter attractor 
\begin{equation}
\quad \Omega _{\phi }=\frac{g}{4(\beta +\mu )^{2}},\quad p=\frac{2}{3}\left(
1+\frac{\beta }{\mu }\right) .
\end{equation}
It is convenient to note that $\beta /\mu $ is a measure of the deviation
from the uncoupled $p_{0}=2/3$ law in MDE: 
\begin{equation}
\frac{\beta }{\mu }=\frac{C}{s}=\frac{3p}{2}-1=\frac{p}{p_{0}}-1=\frac{ 
\delta \rho }{p}.
\end{equation}
We give also the relation between the parameters $(\beta ,\mu )$ and the
observables $\left( \Omega _{\phi },p\right) :$ 
\begin{eqnarray}
g &=&\frac{18p\Omega _{\phi }}{p_{0}-p(1-\Omega _{\phi })},  \nonumber \\
\beta  &=&\frac{(g-18)}{2}\left( \frac{\Omega _{\phi }}{g}\right)
^{1/2},\quad \mu =\frac{1}{2}\left[ \left( \frac{g}{\Omega _{\phi }}\right)
^{1/2}-2\beta \right] .
\end{eqnarray}
For small $\frac{\delta \rho }{p}$ we have 
\begin{equation}
C\simeq \kappa \sqrt{3}\Omega _{\phi }^{-1/2}\frac{\delta \rho }{p}.
\label{bound}
\end{equation}

Since the slope and the matter content in the MDE depend on the model
parameters, the equivalence epoch (subscript $e$) also depends on them. It
is easy to see that the following relation holds 
\begin{equation}
a_{e}^{4-3p_{0}/p}=\frac{\rho _{\gamma 0}}{\rho _{m0}}.
\end{equation}
Clearly, the equivalence occurs earlier with respect to the uncoupled case
if $p<2/3$ (that is, $C/s<0$), later if $p>2/3$ (that is, $C/s>0$).

We will make often use of the fact that on the attractor in the RDE
(subscript R) and in the MDE (subscript M) we have 
\begin{equation}
\phi =\alpha _{M,R}\log a,
\end{equation}
where 
\begin{equation}
\alpha _{R}=-\frac{4}{s},\quad \alpha _{M}=-\frac{3}{s+C}.
\end{equation}
Finally, it is useful to note that 
\begin{equation}
C\alpha _{M}=-\frac{3C}{s+C}\simeq -3\frac{\delta p}{p}
\end{equation}
(the latter is valid for $\frac{\delta p}{p}\ll 1$).

\section{Perturbations}

We now proceed to study the evolution of the perturbations in the coupled $ 
\phi $CDM theory. This involves the following tasks: 1) calculate the linear
perturbation equations (we choose the synchronous gauge for the perturbed
metric) for the coupled system of baryons (subscript $b$), CDM ($c$),
radiation ($\gamma $), scalar field ($\phi $), massless neutrinos ($\nu $);
2) establish initial conditions (we adopt adiabatic initial conditions); 3)
evolve the equations from deep into the radiation era and outside the
horizon down to the present; 4) calculate the radiation fluctuations on the
microwave background and the matter power spectrum at the present; 5)
compare with observations.

Let us identify the effects of adding a
 scalar field  to standard CDM.
The  field component clearly induces two main consequences for as
concerns the perturbation equations: delays the epoch of equivalence,
because the matter density at the present is smaller than without scalar
field, and changes the perturbation
equations. The first effect induces a turn-over of the power spectrum at
larger scales, just as in the case of an open universe, or a model with a
large cosmological constant, so that the power spectrum normalized to COBE
has less power on small scales, as observed. The modification to the
perturbation equations goes in the same direction: the evolution in the MDE
for sub-horizon modes is suppressed with respect to standard CDM, as we will
see below. The evolution equations in the other cases (super-horizon modes,
RDE) give the same behavior as for the pure CDM . The net result is that FJ
find that $\Omega _{\phi }\simeq 0.1$ gives a good fit to observations,
comparable or superior to MDM or $\Lambda $CDM$.$

When we insert the coupling, the two effects above mentioned are again the
dominant ones. But now, the consequences of the coupling can be in either
directions, that is, the equivalence epoch can be delayed or anticipated,
and the perturbations can be either suppressed or enhanced with respect to
the uncoupled case, although not by a large factor. To understand this
effects we first discuss analytically the perturbation equations. Following
the discussion in FJ, we simplify the problem by reducing the system to
three components, CDM, scalar field, and radiation. The notation is

\begin{equation}
\delta =\delta \rho /\rho ,\quad \varphi =\delta \phi ,\quad v_{i}=\delta
u_{i},\quad ik^{i}v_{i}=\theta .
\end{equation}
where $u_{i}$ is the comoving velocity. The perturbation equations in
synchronous gauge are:

Scalar field equation: 
\begin{equation}
\ddot{\varphi}+2H\dot{\varphi}+k^{2}\varphi +a^{2}U_{,\phi \phi }\varphi + 
\frac{1}{2}\dot{h}\dot{\phi}=Ca^{2}\rho _{m}\Omega _{c}\delta _{c},
\end{equation}
CDM: 
\begin{eqnarray}
\dot{\delta}_{c} &=&-\theta _{c}-\frac{1}{2}\dot{h}-C\dot{\varphi},
\label{decdm} \\
\dot{\theta}_{c} &=&-H\theta _{c}+C(k^{2}\varphi +\dot{\phi}\theta _{c}).
\end{eqnarray}
Radiation: 
\begin{eqnarray}
\dot{\delta}_{\gamma } &=&-\frac{4}{3}\theta _{\gamma }-\frac{2}{3}\dot{h},
\\
\dot{\theta}_{\gamma } &=&\frac{k^{2}}{4}\delta _{\gamma }.
\end{eqnarray}
Energy-momentum tensor: 
\begin{eqnarray}
a^{2}\delta T_{0}^{0} &=&a^{2}\left( \delta _{c}\rho _{c}+\delta _{\gamma
}\rho _{\gamma }\right) +\dot{\phi}\dot{\varphi}+a^{2}U_{,}\varphi  \\
\frac{a^{2}}{k^{2}}ik^{i}\delta T_{i}^{0} &=&\frac{a^{2}}{k^{2}}w_{\gamma
}\theta _{\gamma }\rho _{\gamma }+\dot{\phi}\varphi  \\
a^{2}\delta T_{i}^{i} &=&-a^{2}\delta _{\gamma }\rho _{\gamma }-3\left( \dot{ 
\phi}\dot{\varphi}-a^{2}U_{,}\varphi \right) .
\end{eqnarray}
Metric: 
\begin{eqnarray}
H\dot{h} &=&2k^{2}\eta +8\pi a^{2}\delta T_{0}^{0} \\
\dot{\eta} &=&4\pi \frac{a^{2}}{k^{2}}ik^{i}\delta T_{i}^{0} \\
\ddot{h} &=&-H\dot{h}-8\pi a^{2}(\delta T_{0}^{0}-\delta T_{i}^{i}).
\label{hdotdot}
\end{eqnarray}

Deriving Eq. (\ref{decdm}) and inserting equation (\ref{hdotdot}) we get 
\begin{equation}
\ddot{\delta}_{c}+H\dot{\delta}_{c}-\frac{3}{2}H^{2}(\Omega _{c}\delta
_{c}+2\Omega _{\gamma }\delta _{\gamma })-8\pi \left( 2\dot{\phi}\dot{\varphi 
}-sa^{2}U\varphi \right) +C(H\dot{\varphi}+k^{2}\varphi +\ddot{\varphi} 
-4H^{2}\Omega _{\gamma }\varphi )=0.  \label{peq}
\end{equation}
The equation for the scalar field becomes (putting $\theta _{c}=0$) 
\begin{equation}
\ddot{\varphi}+2H\dot{\varphi}+k^{2}\varphi +s^{2}a^{2}U\varphi -\dot{\delta} 
_{c}\dot{\phi}=C\left( \frac{3H^{2}}{8\pi }\Omega _{c}\delta _{c}+\dot{\phi} 
\dot{\varphi}\right) .
\end{equation}
Finally, the radiation equation is 
\begin{equation}
\ddot{\delta}_{\gamma }+\frac{k^{2}}{3}\delta _{\gamma }-\frac{4}{3}\left( 
\ddot{\delta}_{c}+\dot{\theta}_{c}+C\ddot{\varphi}\right) =0.
\end{equation}
The adiabatic initial condition gives now, putting for the initial value of
the scalar field $\varphi =\chi \delta _{c}$ ($\chi $ is determined below), 
\begin{equation}
\delta _{\gamma }=\frac{4}{3}\delta _{c}\left( 1+C\chi \right) .
\end{equation}
In the large scale limit, $k^{2}\rightarrow 0$, and in RDE, where $H=\tau
^{-1}$ and $\Omega _{c}\rightarrow 0,$ and assuming the adiabatic condition,
the system reduces to 
\begin{eqnarray}
\ddot{\delta}_{c}+\tau ^{-1}\dot{\delta}_{c}-4\tau ^{-2}\delta _{c}\Omega
_{\gamma }-8\pi \left( 2\dot{\phi}\dot{\varphi}-sa^{2}U\varphi \right)
+C(\tau ^{-1}\dot{\varphi}+\ddot{\varphi}-4\tau ^{-2}\Omega _{\gamma
}\varphi ) &=&0,  \nonumber \\
\ddot{\varphi}+2\tau ^{-1}\dot{\varphi}+s^{2}a^{2}U\varphi -\dot{\delta}_{c} 
\dot{\phi}-C\dot{\phi}\dot{\varphi} &=&0.
\end{eqnarray}
Inserting the RDE attractor solution for $\phi $, we obtain that the growing
mode both for $\delta $ and $\varphi $ goes as $\tau ^{2}.$ Therefore, the
super-horizon perturbations in RDE grow similarly in CDM, in $\phi $CDM$,$
and in coupled $\phi $CDM. Moreover, we have that, initially, 
\begin{equation}
\varphi =-\frac{4}{5s}\delta _{c}\left( 1+\frac{4C}{5s}\right) ^{-1}\equiv
\chi \delta _{c}.
\end{equation}
Therefore, the initial condition for the CDM density fluctuations on the
attractor in the RDE is 
\begin{equation}
\delta _{c}=-\frac{1}{2}\frac{h}{1+C\chi }.
\end{equation}

Now we consider the super-horizon modes in MDE. The equations are now 
\begin{eqnarray}
\ddot{\delta}_{c}+H\dot{\delta}_{c}-\frac{3}{2}H^{2}\Omega _{c}\delta
_{c}-8\pi \left( 2\dot{\phi}\dot{\varphi}-sa^{2}U\varphi \right) +C(H\dot{
\varphi}+\ddot{\varphi}) &=&0,  \nonumber \\
\ddot{\varphi}+2H\dot{\varphi}+s^{2}a^{2}U\varphi -\dot{\delta}_{c}\dot{\phi}
-C\left( \frac{3H^{2}}{8\pi }\Omega _{c}\delta _{c}+\dot{\phi}\dot{\varphi}
\right) &=&0.
\end{eqnarray}
The growing mode is again $\tau ^{2}$, that is, there is no difference with
respect to the standard case.

In the sub-horizon regime, neglecting the gravitational feed-back, we have
in RDE

\begin{eqnarray}
\ddot{\delta}_{c}+H\dot{\delta}_{c}-4H^{2}\Omega _{\gamma }\delta _{\gamma
}-8\pi \left( 2\dot{\phi}\dot{\varphi}-sa^{2}U\varphi \right) +C(H\dot{ 
\varphi}+k^{2}\varphi +\ddot{\varphi}-4H^{2}\Omega _{\gamma }\varphi ) &=&0
\label{peq2} \\
\ddot{\varphi}+H(2-C\alpha )\dot{\varphi}+k^{2}\varphi  &=&0 \\
\ddot{\delta}_{\gamma }+\frac{k^{2}}{3}\delta _{\gamma } &=&0.
\end{eqnarray}
The oscillating behavior of $\varphi $ and of $\delta _{\gamma }$ gives a
negligible influence on $\delta _{c}$, so that 
\begin{equation}
\ddot{\delta}_{c}+H\dot{\delta}_{c}=0.
\end{equation}
which gives $\delta _{c}=const,\log \tau $, once again with no difference
with respect to standard CDM.

We finally come to the regime where the new physics makes the difference. In
the sub-horizon MDE regime, neglecting again the gravitational feed-back, we
have 
\begin{eqnarray}
\ddot{\delta}_{c}+H\dot{\delta}_{c}-\frac{3}{2}H^{2}\Omega _{c}\delta
_{c}-8\pi \left( 2\dot{\phi}\dot{\varphi}-sa^{2}U\varphi \right) +C\left[ -H 
\dot{\varphi}+C\left( \frac{3H^{2}}{8\pi }\Omega _{c}\delta _{c}+\dot{\phi} 
\dot{\varphi}\right) \right]  &=&0 \\
\ddot{\varphi}+2H\dot{\varphi}+k^{2}\varphi -C\left( \frac{3H^{2}}{8\pi } 
\Omega _{c}\delta _{c}+\dot{\phi}\dot{\varphi}\right)  &=&0.
\end{eqnarray}
Neglecting the oscillating behavior of $\varphi $ we obtain 
\begin{equation}
\ddot{\delta}_{c}+H(1+C\alpha _{M})\dot{\delta}_{c}-\frac{3}{2}H^{2}\Omega
_{c}\delta _{c}(1-\frac{C^{2}}{4\pi })=0.  \label{basic}
\end{equation}
Inserting the trial solution $\delta _{c}=Ba^{m}$ we obtain two solutions
for $m$: {\Large \ } 
\begin{equation}
m_{\pm }=\frac{1-p}{2p}\left\{ -1\pm \left[ 1+F(\Omega _{\phi },p)\right]
^{1/2}\right\} ,
\end{equation}
where{\Large \ } {\Large \ } 
\begin{equation}
F(\Omega _{\phi },p)=\frac{6p(1-\Omega _{\phi })\left[ -8+26p+3(\Omega
_{\phi }-7)p^{2}\right] }{(p-1)^{2}\left[ 2+3p(\Omega _{\phi }-1)\right] }.
\label{mpm}\end{equation}
For{\Large \ }$p=2/3${\Large \ }this reduces to the form found in FJ{\Large  
\ } 
\begin{equation}
m_{\pm }=\frac{1}{4}\left( -1\pm \sqrt{25-24\Omega _{\phi }}\right) .
\end{equation}
In Fig. 3 we show the contour plot of $m_{+}(\Omega _{\phi },p)$.
 This figure is crucial for the
understanding of the perturbation evolution, so we discuss it at some
length. First, we observe that for {\it all} values of $\Omega _{\phi },p$
there is suppression with respect to CDM: the slope is in fact always less
than 1 and, for $p\simeq 2/3$, the slope is smaller for larger $\Omega
_{\phi }$. Second, we notice the unexpected fact that the value $p=2/3$ is
close to the maximum for all values of $\Omega _{\phi }$, and closest for
small $\Omega _{\phi }$. For $\Omega _{\phi }=0.1$, for instance, the
maximum is at $p=0.672$, while for $\Omega _{\phi }=0.6$ it is at $p=0.710$.
This implies immediately that the coupling does not enhance much the
fluctuation growth with respect to the uncoupled case, while it can sensibly
reduce it further as long as $p$ is far from $2/3.$ Third, there is only a
finite range of $p$, almost centered around $2/3$, for which real values of $ 
m_{\pm }$ exist. Beyond that range, the power-law solutions of Eq. (\ref
{basic}) are replaced by oscillating solutions $\cos \left( \log a\right) $,
in which the restoring force is the coupling interaction.

Let us then summarize the asymptotic evolution of the fluctuations in the
coupled model. There are two relevant cases. If $p>2/3$, the equivalence
epoch occurs later than in the uncoupled case. Then, smaller wavenumbers
reenter during the RDE than in the uncoupled case, and therefore there is
extra suppression at these scales. Then, in the subsequent MDE regime, the
modes are further suppressed with respect to the uncoupled case, unless $p$
is close to $2/3$. The transfer function will be then more steeply declining
with respect to the uncoupled case. If $p<2/3$, on the other hand, the
equivalence occurs earlier, and the scales smaller than $2\pi \tau _{e}/a_{e}
$ are less suppressed. At the same time, the MDE regime induces again a
slower fluctuation growth, so that there is an intermediate region of
wavenumbers with a depleted transfer function, and a large wavenumber region
with an enhanced transfer function. Fig. 4 displays some of these features.

The only important difference that arises when the baryons are added is in
the tight coupling approximation. Referring to the notation used in Ma \&
Bertschinger (1995), we have the two equations 
\begin{eqnarray}
\dot{\theta}_{\gamma } &=&k^{2}\left( \frac{1}{4}\delta _{\gamma }-\sigma
_{\gamma }\right) +\frac{1}{\tau _{c}}\left( \theta _{b}-\theta _{\gamma
}\right)   \label{the1} \\
\dot{\theta}_{b} &=&-H\theta _{b}+c_{s}^{2}k^{2}\delta _{b}-\frac{R}{\tau
_{c}}\left( \theta _{b}-\theta _{\gamma }\right) +C(k^{2}\varphi +\dot{\phi} 
\theta _{b}).  \label{the2}
\end{eqnarray}
The slip equation $\dot{\theta}_{b}-\dot{\theta}_{\gamma }$ in the tight
coupling approximation can be derived exactly as in Ma \& Bertschinger
(1995), taking into account that now (here $n_{e}$ is the electron density
and $\sigma _{T}$ the Thomson cross section)  
\begin{eqnarray}
\tau _{c} &=&(an_{e}\sigma _{T})^{-1}\sim a^{-2}e^{C\phi },\quad \dot{\tau} 
_{c}=\left( 2H+C\dot{\phi}\right) \tau _{c}  \nonumber \\
R &=&\frac{4\rho _{\gamma }}{3\rho _{b}},\quad \dot{R}=\left( H-C\dot{\phi} 
\right) R.
\end{eqnarray}
To second order in $\tau _{c}$ we obtain that the slip between baryons and
photons is 
\begin{eqnarray}
\dot{\theta}_{b}-\dot{\theta}_{\gamma } &=&\frac{2(H-C\dot{\phi})R}{1+R} 
\left( \theta _{b}-\theta _{\gamma }\right)   \nonumber \\
&&+\frac{\tau _{c}}{1+R}[-\frac{\ddot{a}}{a}\theta _{b}-\frac{1}{2}\left( H+ 
\frac{1}{2}C\dot{\phi}\right) k^{2}\delta _{\gamma }+k^{2}\left( c_{s}^{2} 
\dot{\delta}_{b}-\frac{1}{4}\dot{\delta}_{\gamma }\right) +  \nonumber \\
&&C\left( Hk^{2}\varphi +H\dot{\phi}\theta _{b}+k^{2}\dot{\varphi}+\ddot{\phi 
}\theta _{b}\right) ].
\end{eqnarray}
The equation for the photons is 
\begin{equation}
\dot{\theta}_{\gamma }=-R^{-1}\left[ \dot{\theta}_{b}+H\theta
_{b}-k^{2}c_{s}^{2}\delta _{b}-C(k^{2}\varphi +\dot{\phi}\theta _{b})\right]
+k^{2}\left( \frac{1}{4}\delta _{\gamma }-\sigma _{\gamma }\right) .
\end{equation}

This concludes the analysis of the asymptotic regimes in the coupled $\phi $
CDM model. The results that will be presented in the next Sections make use
of the full machinery of the Boltzmann code, as implemented in the CMBFAST
code of Seljak and Zaldarriaga (1996), opportunely modified to take into
account the coupled scalar field (including the transient from the RDE
attractor to the MDE one). The equations are essentially the same as in FJ,
with the new terms due to the coupling as detailed above.  We tested the
code with the results of FJ when $p=2/3$, and we also checked our results
with the asymptotics found above.

\section{Comparison with observations: cosmic microwave background}

The main effect of the coupling on the cosmic microwave background is on the
location and amplitude of the acoustic peaks. The location of the peak is
related to the size of the sound horizon at decoupling (subscript $d$).
Since the photon-baryon fluid has sound velocity 
\begin{equation}
c_{spb}^{2}=\frac{1}{3}r,\quad r\equiv \frac{R}{w_{c}^{\prime }+R},
\end{equation}
where $w_{c}^{\prime }=1+C\alpha /3$, the sound horizon is 
\begin{equation}
r_{s}=\int_{0}^{\tau _{d}}\frac{d\tau }{\left( 3r\right) ^{1/2}} 
=\int_{0}^{a_{d}}\frac{da}{\left( 3H^{2}a^{2}r\right) ^{1/2}}.
\end{equation}
This expression can be simplified as follows. First, we put ourselves in the
case $a_{e}\ll a_{d}\simeq 10^{-3}$ and neglect the RDE stage altogether. In
MDE we have 
\begin{equation}
H^{2}a^{2}\simeq H_{0}^{2}ae^{-C\phi }.
\end{equation}
Then we can write, remembering that on the attractor $e^{-C\phi
}=a^{-C\alpha },$ and defining the standard sound horizon $ 
r_{s0}=2a_{d}^{1/2}H_{0}^{-1}/\sqrt{3}$ 
\begin{equation}
r_{s}=\frac{r_{s0}}{a_{\ast }^{1/2}}\int_{0}^{a_{d}}\frac{da}{2\left(
a^{1-C\alpha }r\right) ^{1/2}}.
\end{equation}
We can further simplify, for $r\simeq 1$, i.e. $R\gg 1$ (which is true at
decoupling) 
\begin{equation}
r_{s}=r_{s0}\frac{a_{d}^{C\alpha /2}}{1+C\alpha },
\end{equation}
and the corresponding peak multipole is, for $\delta p/p\ll 1$ 
\begin{equation}
\ell _{peak}\simeq \frac{2\pi }{r_{s}H_{0}}=\ell _{0}\left( 1+C\alpha
\right) a_{d}^{-C\alpha /2}=\ell _{0}\left( 1-3\frac{\delta p}{p}\right)
a_{d}^{1.5\delta p/p},
\end{equation}
where the standard peak multipole is 
\begin{equation}
\ell _{0}=\frac{2\pi }{r_{s0}H_{0}}\simeq 200.
\end{equation}
The qualitative behavior is clear: for $p<2/3$ there is a larger $\ell
_{peak}$ than in the uncoupled model, for $p>2/3$ a smaller $\ell _{peak}.$
For instance, for $p=0.65$ we expect $\ell _{peak}\simeq 250,$ in agreement
with the numerical results.

We calculated the $C_{\ell }$ spectrum for several coupled $\phi $CDM
models, parametrized by the two observables $\Omega _{\phi },p$. The range
of values we explore, in this and in the next Section, is 
\begin{equation}
\Omega _{\phi }=0.05-0.2,\quad p=0.65-0.70.
\end{equation}
The values of the other relevant parameters are fixed as follows 
\begin{equation}
h=0.7,\quad \Omega _{b}=0.04,\quad \Omega _{\Lambda }=0,\quad n=1.
\end{equation}
In Fig. 5 we display the multipole spectra. As anticipated, the acoustic
peaks move to larger multipoles as $p$ decreases.

There are two other effects worth discussing: the amplitude of the acoustic
peaks and the slope of the multipole spectrum at small $\ell $. The
amplitude of the peak is depressed as $\delta p/p$ increases, save for
values close to 2/3, since the matter fluctuations that drive the radiation
peaks are suppressed, as shown above. The small $\ell $ region is dominated
by the Sachs-Wolfe (SW) effect. As well known, the integrated SW (ISW)
effect in flat space vanishes only if the fluctuations grow as $a$, which is
not the case here. The ISW then adds at small multipoles and tilts the $ 
C_{\ell }$ spectrum. Moreover, the overall normalization now takes into
account the ISW power, and as a consequence the normalization for the
perturbation at decoupling time is reduced. This effect shows also in the
final amplitude of the power spectrum.

Deriving precise constraints from the whole set of observations on the CMB
requires considerable detail in the statistical procedure, beyond the scope
of this paper. Here we content ourselves to derive rough limits on the
parameters. It is probably safe to state that current observations rule out
values $p<0.63$ or $p>0.72$, although the present level of errors does not
permit to attach a strong significance to such bounds. Future precision
observations around the first peaks are likely to constrain $p$ to two
decimal digits. As already found by FJ, on the other hand, the microwave sky
does not impose strong constraints on $\Omega _{\phi }$, since this
parameter influences mainly the fluctuation growth, and thus the absolute
normalization. To constrain it, we have to evaluate the present power
spectrum of the fluctuations.

\section{Comparison with observations: power spectrum}

The analytical expression (\ref{mpm}) 
for the fluctuation growth exponent found in
Section 3 is a clear guide to the results of this Section. As anticipated,
the coupling introduces an extra suppression for the scales that enter the
horizon in the MDE. The suppression is 
larger for models with high $\Omega _{\phi }$ and high $\mid \delta p/p\mid $
. A small suppression factor, as well known, helps to bring the standard CDM
model into agreement with observations. FJ found that the best uncoupled
models have $\Omega _{\phi }=0.1$; here we see that the coupling allows also
models with smaller $\Omega _{\phi }$, but $\delta p/p\neq 0,$ to meet the
observations. This can be helpful to reduce the constraints from
nucleosynthesis, which, in some restrictive analysis, require $\Omega _{\phi
}<0.1$.

In Fig. 6 we report the power spectra $\Delta ^{2}(k)=k^{3}P(k)/(2\pi ^{2})$
normalized to COBE, compared to the data as compiled and corrected for
redshift and non-linear distortions by Peacock \& Dodds (1994). 
 For a quantitative comparison, we
plot in Fig. 7 the contour plot of $\sigma _{8}(\Omega _{\phi },p)$, the
number density variance in 8 $Mpc$/h spheres. The
models with a value of $\sigma _{8}$ larger than 0.5, as required by cluster
abundance (White, Efstathiou \& Frenk 1993, Viana \& Liddle 1996, Girardi et
al. 1998), have $\Omega _{\phi }<0.15$ and at most a small deviations from $ 
p=2/3$. The suppression of $\sigma _{8}$ with respect to COBE-normalized
standard CDM is due both to the growth suppression in MDE and to the fact
that now the COBE normalization includes the ISW effect.

For as concerns the shape of the spectrum, the comparison with the galaxy
data is uncertain due to biasing. Assuming a scale-independent bias between
matter and galaxies, we can quantify the agreement with the data by
evaluating the $\chi ^{2}$ of the ratio between the theoretical and the
galaxy spectrum, $R_{i}=P_{G}(k_{i})/P_{T}(k_{i}),$ that is by evaluating 
\begin{equation}
\chi ^{2}=\sum_{i}\left[ R_{i}-\hat{R}\right] ^{2}/\sigma ^{2}(k_{i}),
\end{equation}
where $\hat{R},\sigma ^{2}(k)$ are the average and variance of $R_{i}$,
neglecting cosmic variance. $\ $The contour plots of $\chi ^{2}(\Omega
_{\phi },p)$ are in Fig. 8. They show, as anticipated, that the models with $ 
p>2/3$ follow better the real data because are more bent at small scales.
The best models among those studied here have $\chi ^{2}\simeq 13$ for $N=$
15 degrees of freedom (16 real data, minus the average $\hat{R}$ estimated
from the data themselves). For instance the model with $\Omega _{\phi
}=0.15,p=0.685$ gives a very good fit, and has $\sigma _{8}\simeq 0.6$ as
required. Notice that we performed the fits without varying all the other
cosmological parameters, which, at least in principle, can be determined by
other observations.

In Fig. 9 we summarize the constraints obtained in this Section, considering
the models which have $\sigma _{8}\in (0.45,0.75)$ and $\chi ^{2}/N<2$. The 
cluster abundance normalization usually quoted for $\Omega _{m}\simeq 1$ is  
$\sigma _{8}\simeq 0.60\pm 0.05$, but this is calculated for standard
models, so that  conservatively  a  larger region has been adopted. Only a
stripe around $p=0.67-0.69$ and $\Omega _{\phi }=0.05-0.2$ appears to be
viable, if the bias is indeed scale-independent. From Eq. (\ref{bound}) we
deduce a limit 
\begin{equation}
0<C\lesssim 1M_{p}^{-1}
\end{equation}
which, although  still far from the limit 0.1$M_{P}^{-1}$
 quoted in Wetterich
(1995) from local measures,  is global and applies even if the scalar field
is not coupled to baryons, as proposed in Damour, Gibbons \& Gundlach
(1990). Future data will certainly tighten the constraint even further.

\section{Conclusion}

In this paper we discussed the perturbations of a coupled scalar field with
exponential potential on the CMB and on the present large scale structure
along an attractor solution. For as concerns the CMB, we found that the
coupling induces a strong effect on the location and amplitude of the
acoustic peaks, due to the variation of the scale factor expansion law.
Future precision measures in the region $\ell >100$ have the potential of
constraining the coupling to two orders of magnitudes better than at the
present (see for instance the discussion in Chen \& Kamionkowsky 1999).

We found that subhorizon perturbations are always more suppressed in MDE
with respect to standard CDM, no matter what the parameters $\Omega _{\phi }$
and $p$ are. Moreover, the suppression increases for $p$ far from the
standard value. The amplitude $\sigma _{8}$ at the present is between 0.5
and 1 only for $\Omega _{\phi }\in (0.05-0.2)$, being smaller for larger $ 
\Omega _{\phi }$, as already found in the uncoupled case by FJ. Adding the
request to fit the galaxy power spectrum shape, the parametric space is
reduced as in Fig. 9. A positive coupling has the advantage to warp the
spectrum to a closer agreement with the data.

The background solution we adopted here is only one of the possible
solutions. An equally interesting  one is to consider a solution heading
toward the inflationary attractor $a$, but still short of it. This would
provide closure density to a $\Omega _{m}\simeq 0.3$ universe, and an
acceleration as recently claimed, although at the price of sensibility to
the initial conditions. Such a model will be investigated in a future work.

\section*{Acknowledgments}

I am indebted to Carlo Baccigalupi, Francesca Perrotta and Michael Joyce for
useful discussions on the topic.

%\newpage

%\section{Figures}

\newpage
\begin{figure*}
\epsfysize 4in
\epsfbox{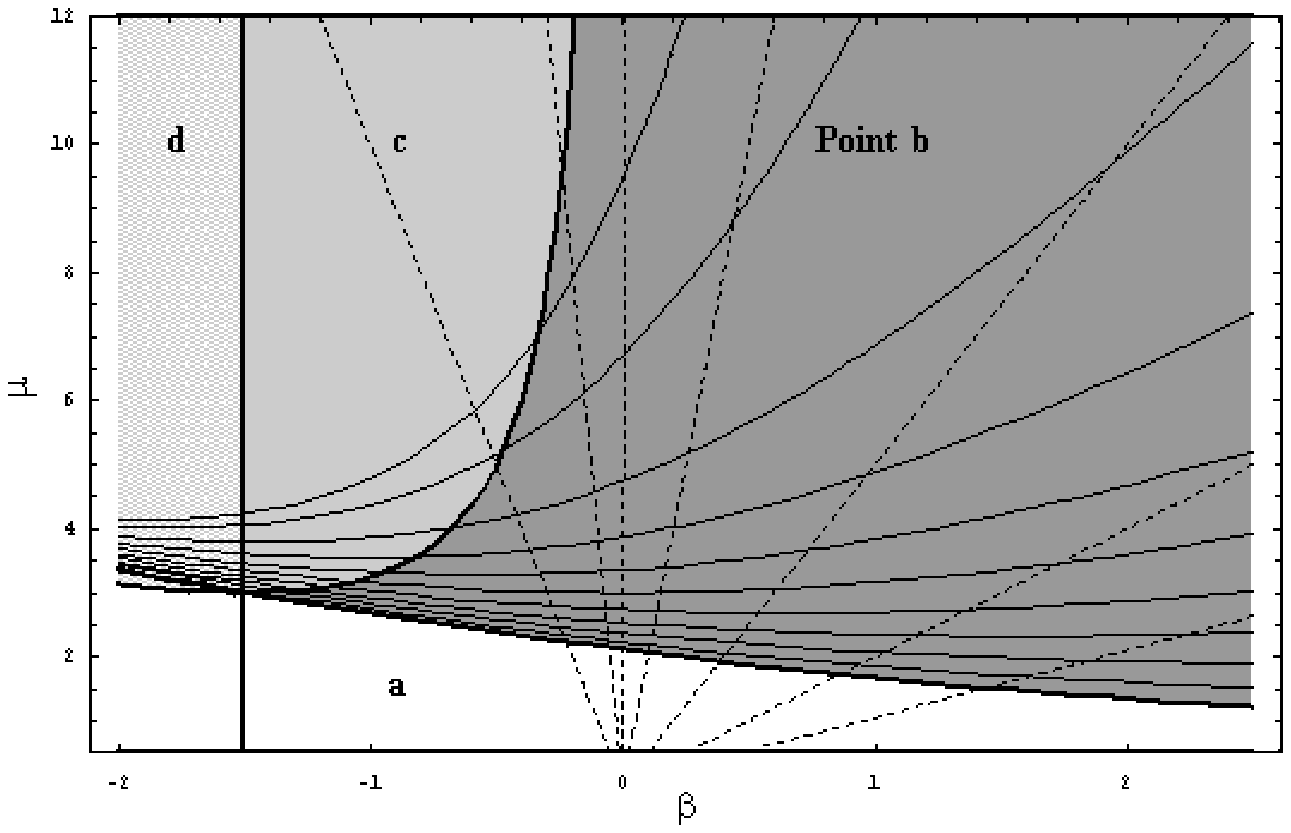}
\caption{ Parameter space for the attractors. In the dark-shaded region
the solution $b$ is an attractor;   in the other regions the attractors
 are the solutions $a,c$ or $d$, as labelled.
 In all the paper we focus on the attractor $b$. The
continuous curves mark values of $\Omega _{\phi }$ equal to
 0.05,0.1, 0.2, 0.3,...,0.9,
top to bottom. The dotted lines are values of $p$ equal to
0.6, 0.65, 2/3, 0.7, 0.8, 1, 1.3, left to right. There exists also a completely
equivalent symmetric region with $\beta \rightarrow -\beta $ and $\mu
\rightarrow -\mu .$}
\end{figure*}

\begin{figure*}
\epsfysize 4in
\epsfbox{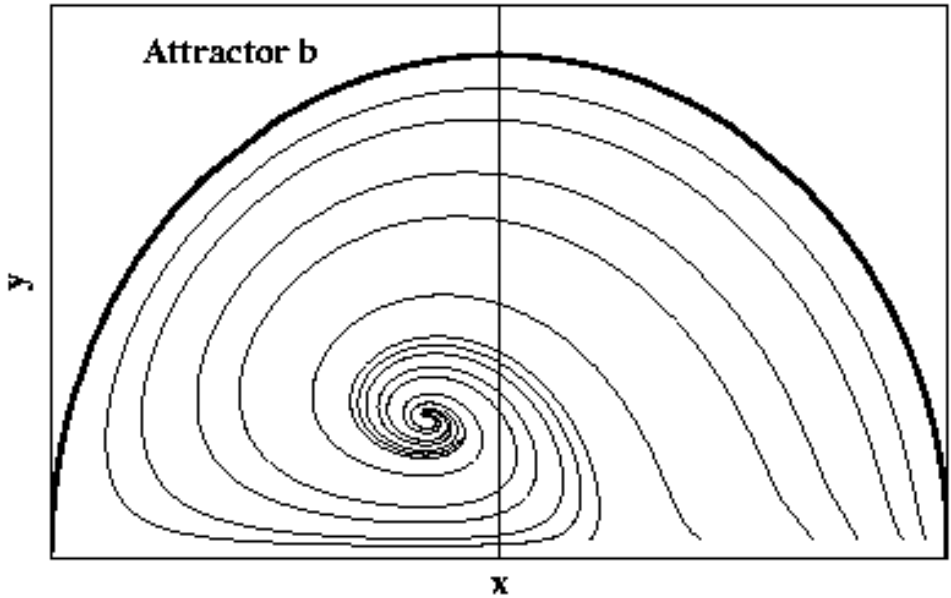}
\caption{
 Phase space corresponding to $\Omega_{\phi}=0.1, p=0.7$, in MDE.}
\end{figure*}

%[100 60 492 500]
\newpage
\begin{figure*}
\epsfysize 6in
\epsfbox{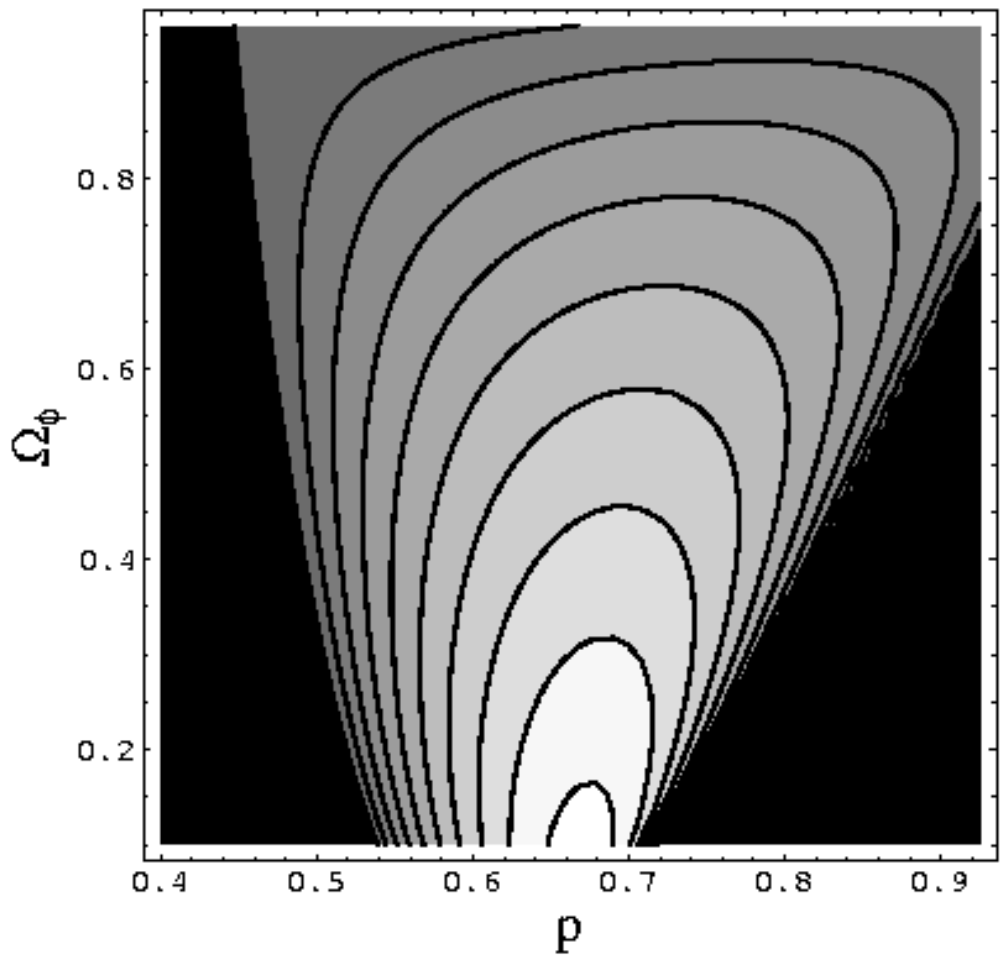}
\caption{ Contour plot of the exponent $m_+$ of the fluctuation growth law $
\delta _{c}\sim a^{m}$ versus $\Omega _{\phi },p$. The contour levels are
for $m=0.9$ (enclosing the white region), down to 0 in steps of 0.1. In the
black region $m$ is complex. Notice that for any given $\Omega _{\phi }$ the
maximum of $m$ is close to $p=2/3$, especially for small $\Omega _{\phi }$.}
\end{figure*}

\newpage
\begin{figure*}
\epsfysize 8in
\epsfbox{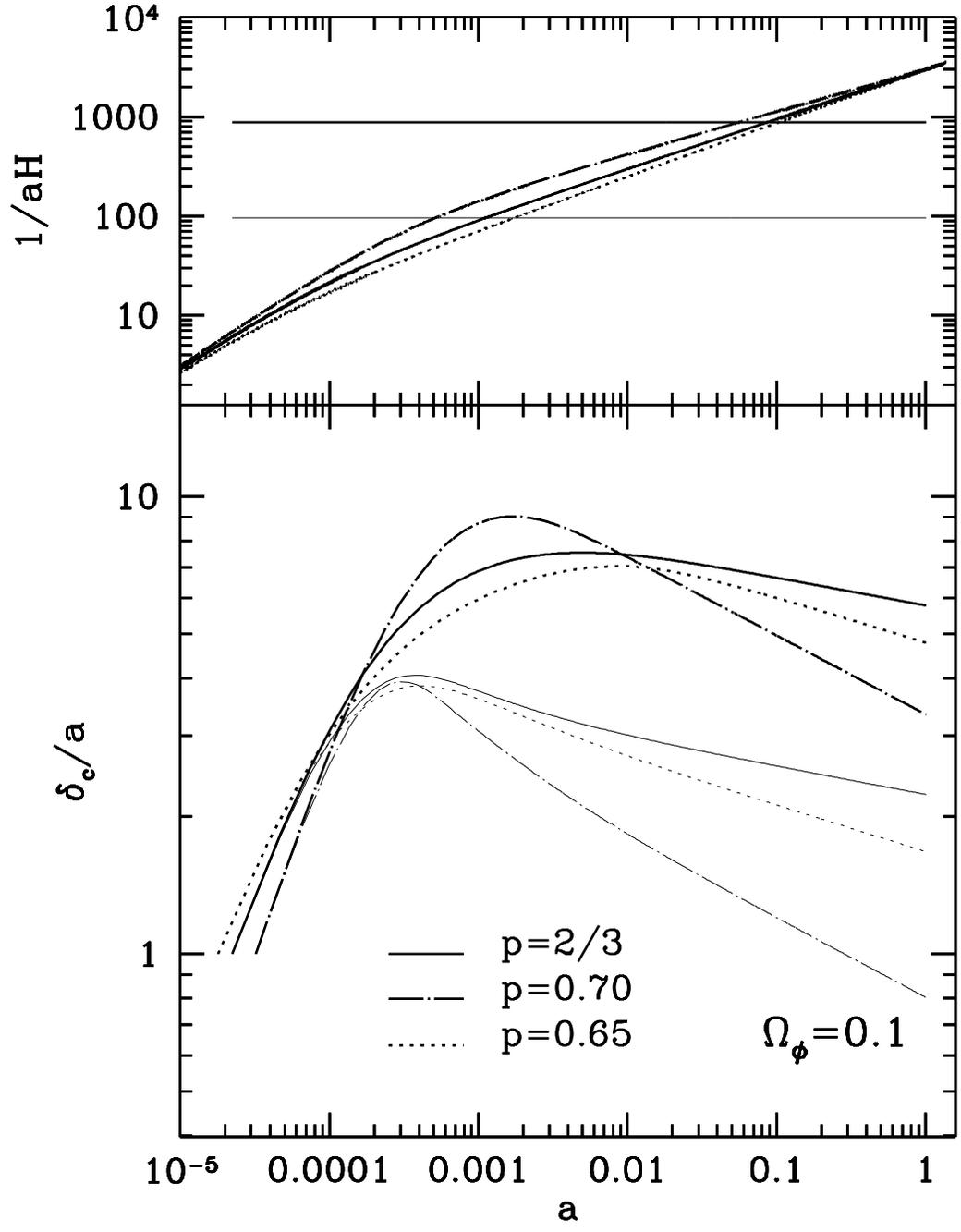}
\caption{ Growth of the dark matter fluctuations for various values of the
coupling and $\Omega_{\phi}=0.1$.
 In the top panel the trend of the horizon length and of two comoving
scale (the horizontal lines) 
show the horizon crossing and the radiation and matter eras. Thick
lines: wavelength $\simeq 900 $Mpc$/h$. Thin lines: 
wavelength $\simeq 100 $Mpc$
/h$.}
\end{figure*}

\newpage

\begin{figure*}
\epsfysize 8in
\centerline{
\epsfbox[-10 4 492 700]{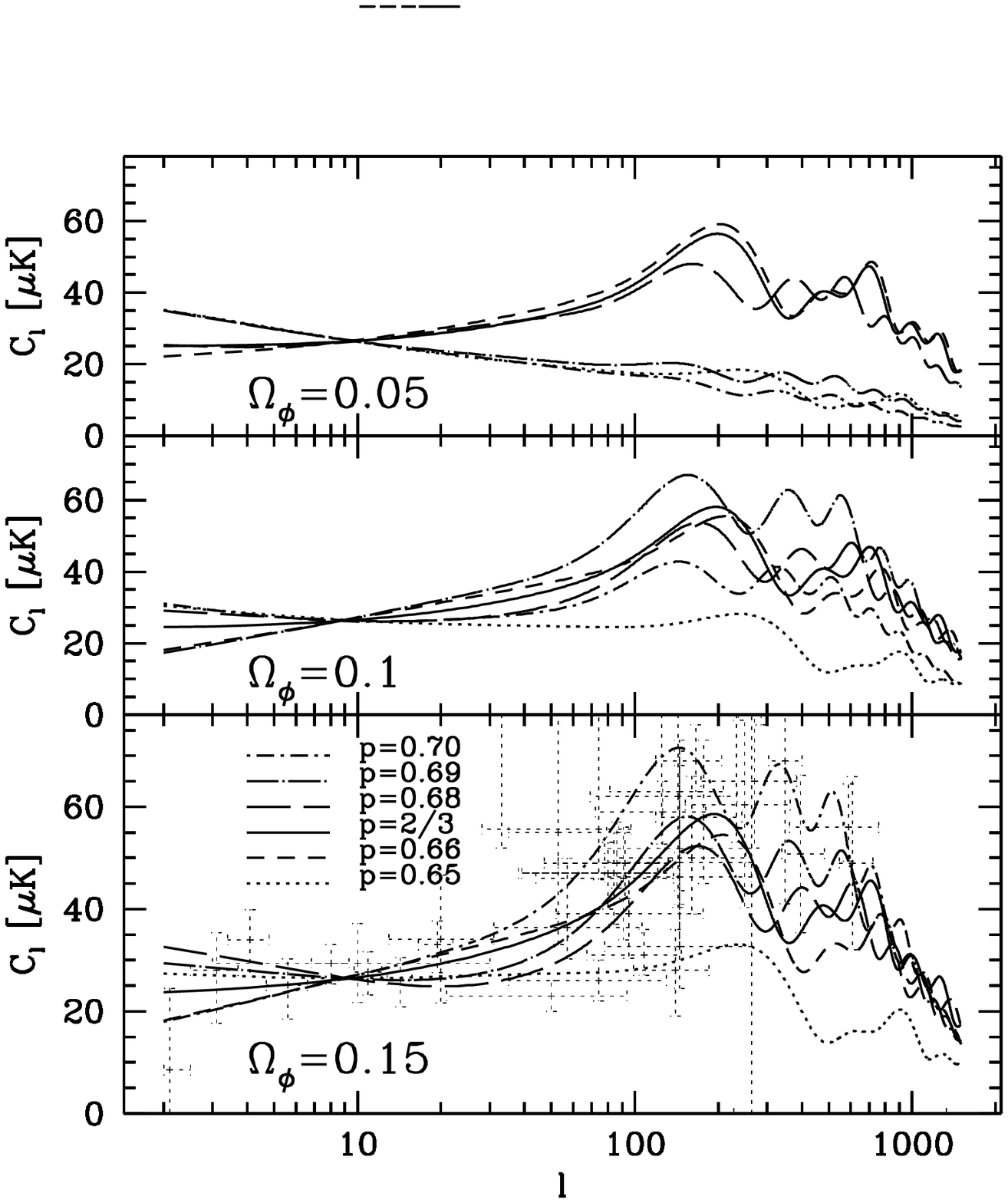}}
\caption{ $C_{\ell }$ spectrum for various models (actually
we plot $[\ell(\ell+1)C_{\ell}/2\pi]^{1/2}$, as
customary). Notice the shift of the
peak location for the different values of $p$, in agreement with the
approximation given in the text. The amplitude decreases for $p\neq 2/3$
(except for values slightly larger than 2/3) and, for a given $p,$ decreases
for smaller $\Omega _{\phi }$ , as expected. \ The data points are from
Tegmark's home page (http:// www.sns.ias.edu/\symbol{126}max). }
\end{figure*}

\newpage
\begin{figure*}
\epsfysize 8in
\epsfbox[-10 4 492 700]{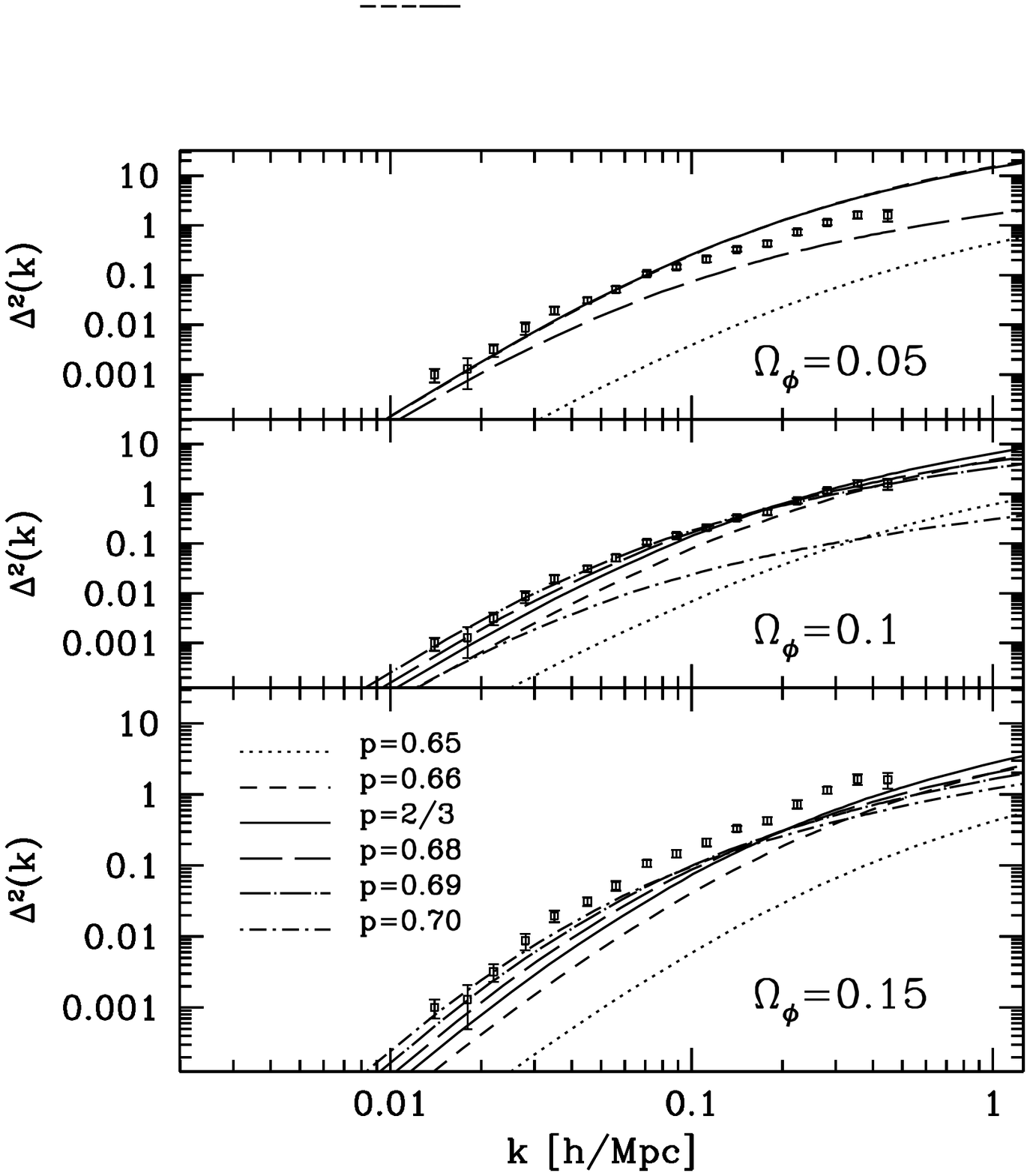}
\caption{ Dark matter power spectra $\Delta ^{2}(k)$ for the same models as
above. Notice again that the power is suppressed as long as $p$ deviates
from the uncoupled law 2/3 (except for values slightly larger than 2/3).
Also notice that models with higher $p$ are more flattened at small scales.}
\end{figure*}

\newpage
\begin{figure*}
\epsfysize 4in
\epsfbox{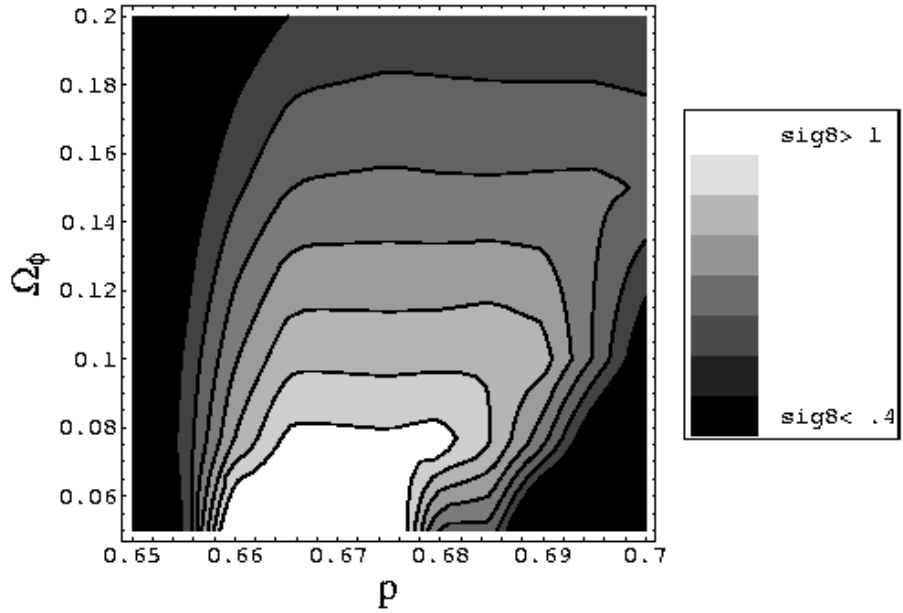}
\caption{ Contour plot of $\sigma _{8}(p,$ $\Omega _{\phi }),$ for $\sigma
_{8}=1, .9, .8, .7, .6, .5, .4$. }
\end{figure*}

\begin{figure*}
\epsfysize 4in
\epsfbox{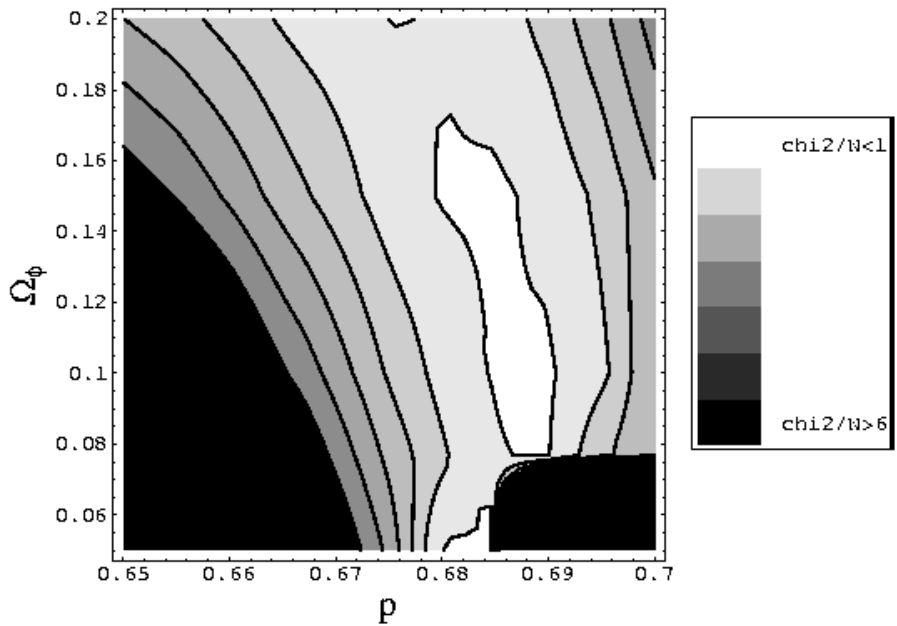}
\caption{ Contour plot of $\chi ^{2}(p,$ $\Omega _{\phi }),$ for $\chi
^{2}/N=1, 2, 3, 4, 5, 6.$}
\end{figure*}

\newpage
\begin{figure*}
\epsfysize 8in
\epsfbox[-10 4 492 700]{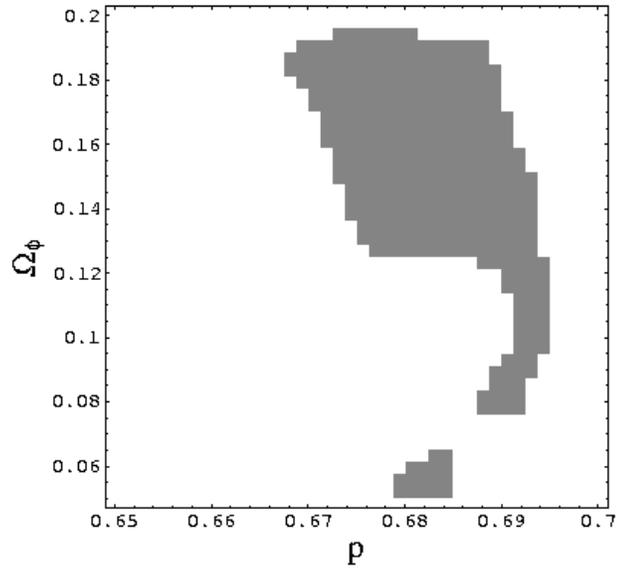}
\caption{The parameters in the shaded region have the correct amplitude and
shape of the power spectrum, i.e. satisfy the constraints $\sigma _{8}\in
(0.45,0.75)$ and $\chi ^{2}/N<2.$}
\end{figure*}

\end{document}